\documentclass[pre,preprint,showpacs,preprintnumbers,amsmath,amssymb]{revtex4}

\usepackage{graphicx}
\usepackage{dcolumn}
\usepackage{bm}

\begin{document}


\title{Universality Aspects of Layering Transitions in Ferromagnetic Blume-Capel Thin Films }
\author{Yusuf Y\"{u}ksel}
\email{yusuf.yuksel@deu.edu.tr}
\author{\"{U}mit Ak{\i}nc{\i}}
\email{umit.akinci@deu.edu.tr}

\affiliation{Department of Physics, Dokuz Eyl\"{u}l University, Kaynaklar Campus, TR-35160 Izmir, Turkey}
\date{\today}
\begin{abstract}
Critical phenomena and universality behavior of ferromagnetic thin films described by a spin-1 Blume-Capel Hamiltonian has been examined for various thickness values ranging from 3 to 40 layers. Using effective field theory, we have found that crystal field interactions significantly affects the critical value of surface to bulk ratio of exchange interactions $R_{c}$ at which the critical temperature becomes independent of film thickness $L$. Moreover, we have extracted the shift exponent $\lambda$ from computed data. Based on the results, we have shown that in the presence of surface exchange enhancement, the system may exhibit a dimensional crossover. We have also found that presence of crystal field interactions does not affect the value of $\lambda$. Hence, a ferromagnetic spin-1/2 thin film is in the same universality class with its spin-1 counterpart.

\end{abstract}

\maketitle

\section{Introduction}\label{intro}
In recent years, influences of surface effects on the magnetic properties of finite systems such as ferromagnetic thin films have attracted a considerable amount of interest \cite{kaneyoshi_review,pleimling}. In addition to remarkable theoretical efforts, preparation of thin films by depositing a magnetic material on a non-magnetic substrate  became experimentally accessible even in the monolayer limit with the development of modern vacuum techniques including molecular beam epitaxy. Due to the presence of free surfaces, magnetic properties of thin films may differ from those of bulk materials. This difference mainly originates from a number of physical phenomena. Namely, the surface atoms have a lower symmetry in comparison with that of the inner atoms \cite{kaneyoshi_00}, and the exchange interactions between the surface atoms may be different from those between the corresponding bulk counterparts. As a consequence of these facts, the surface may exhibit an ordered phase even if the bulk itself is disordered which has already been experimentally observed \cite{ran,polak,tang}.

In this context, an extraordinary case is defined as the transition at which the surface becomes disordered at a particular temperature $T_{c}^{s}$ which is larger than the bulk transition temperature $T_{c}^{b}$. From the academic point of view, due to the fact that many thin films such as the $\mathrm{Fe/Ag(100)}$ system \cite{qiu} exhibit a strong uniaxial anisotropy, phase transition characteristics of thin ferromagnetic films are often modeled by several extensions of an Ising type spin Hamiltonian \cite{strandburg}. It is theoretically predicted that there exists a critical value of surface to bulk ratio of exchange interactions $R_{c}$ above which the surface effects are dominant and the transition temperature of the entire film is determined by the surface magnetization whereas below $R_{c}$, the transition characteristics of the film are governed by the bulk magnetization. The critical value $R_{c}$ itself is called as the special point, and the numerical value of this point has been examined within various theoretical techniques for spin-1/2 case \cite{aguilera,binder_0,kaneyoshi_1,burkhardt,landau_0,sarmento,neto,akinci_2}. Among these works, within the framework of effective field theory (EFT),  Sarmento \textit{et al}. \cite{sarmento} clarified that a transverse field in the surface layer causes the critical value of the surface exchange enhancement $R_{c}$ to move to a higher value whereas the presence of a bulk transverse field causes $R_{c}$ to decrease to a lower value.

The problem has also been handled for higher spins using a number of techniques \cite{zhong,jia,saber,tucker,bahmad,saber_2,zaim,kaneyoshi_balcerzak}. For instance, the phase transition properties of a diluted Blume-Capel ferromagnetic film with $S=1$ in a transverse field have been investigated by EFT \cite{jia}. Using the same method, phase diagrams, the layer longitudinal magnetizations and quadrupolar moments of a spin-1 film have been examined as functions of the ratio of the surface exchange interactions to the bulk ones, transverse fields, and film thickness \cite{saber,saber_2}. In addition, using extensive Monte Carlo (MC) simulations, the effect of surface exchange enhancement on ultrathin spin-1 films has been studied by Tucker \cite{tucker}, and it was concluded that the $R_{c}$ value is spin dependent. However, in a recent work \cite{zaim}, using MC simulations, the influence of crystal-field interaction (or single ion anisotropy) on the critical behavior of a magnetic spin-1 film has been studied, and it has been argued that $R_{c}$ is independent of the crystal-field interaction. It is clear that this latter result apparently conflicts with \cite{tucker}, and according to us it deserves particular attention. Moreover, we see from above discussions that the theoretical studies based on EFT are completely focused on the effect of surface and bulk transverse fields in the absence of crystal fields.

On the other hand, theoretical and experimental investigations are also focused on the finite size shift of the critical temperature of the film as a function of its thickness which is characterized by a shift exponent $\lambda$. A number of experimental studies have been devoted to determine the value of $\lambda$ for various thin film samples, and it has been concluded that the shift exponent extends from $\sim1$ to $3.15$ \cite{farle,elmers,henkel,ballentine}. Since the exponent $\lambda$ is directly related on the bulk correlation length exponent as $\nu_{b}=1/\lambda$ \cite{barber}, within the accuracy of Ising-type models, it can be mentioned that a sample of thin film for which the exponent $\lambda$ is close to unity exhibits a two dimensional character whereas as the value of the exponent becomes larger than unity then the system shows a three dimensional character. Theoretically, the exponent $\lambda$ has been extracted for some certain models with a wide variety of techniques. For instance, using the high temperature series expansion (HTSE) method, it has been shown that the estimated value of $\lambda$ for ferromagnetic Ising \cite{allan,capehard} and Heisenberg \cite{ritchie} thin films severely depends on whether a periodic or free boundary condition was considered in the surface. This result has also been verified within the resolution of MC simulations \cite{binder_thin,kitatani,takamoto,laosiritaworn}. Moreover, the value of the extracted exponent is also very sensitive to the lattice geometry \cite{masrour}.

Under certain circumstances, universal behavior of a thin film system may experience a dimensional crossover. Such a phenomenon has been experimentally observed as the film thickness is varied in ultrathin $\mathrm{Ni}(111)$ films on $\mathrm{W}(110)$ \cite{li_baberschke}, and epitaxial thin films of Co, Ni, and their alloys grown on $\mathrm{Cu}(100)$ and $\mathrm{Cu}(111)$ \cite{huang}. Previous MC simulations \cite{binder_0} also predict that the exponent $\lambda$ may vary continuously with surface exchange $J_{s}$ in the range $R<R_{c}$ which also indicates the occurrence of a dimensional crossover between the surface value and bulk value.

As seen in the above discussions, there are some important questions which remain to be answered regarding the phase transition characteristics of thin ferromagnetic films. For example, the controversy on the crystal field dependence of special point $R_{c}$ should be resolved. In addition, it is not clear whether the presence of a crystal field interaction in the model affects the universality behavior of the system or not. In order to clarify these issues, we intent to study the spin-1 ferromagnetic thin film model in the presence of crystal field interactions within the EFT framework \cite{honmura}. Despite its mathematical simplicity, EFT systematically includes the single-site correlations in the calculations, hence the obtained results are expected to be more accurate than those obtained by conventional mean field theory (MFT).

The organization of the paper is as follows: In Sec. \ref{formulation} we briefly present the formulations. The results and discussions are presented in Sec. \ref{results}, and finally Sec. \ref{conclude} contains our conclusions.

\section{Formulation}\label{formulation}
We consider a ferromagnetic thin film with thickness $L$ described by conventional Blume-Capel Hamiltonian \cite{blume_capel}
\begin{equation}\label{eq1}
\mathcal{H}=-\sum_{<ij>}J_{ij}S_{i}S_{j}-D\sum_{i}(S_{i})^{2},
\end{equation}
where $J_{ij}=J_{s}$ if the lattice sites $i$ and $j$ belong to one of the two surfaces of the film, otherwise we have $J_{ij}=J_{b}$ where $J_{s}$ and $J_{b}$ denote the ferromagnetic surface and bulk exchange interactions, respectively. The first term in Eq. (\ref{eq1}) is a summation over the nearest-neighbor spins with $S_{i}=\pm1,0$ and the second term represents the crystal field energy on the lattice.

The magnetizations and quadrupolar moments (i.e.  $m_{i}=\langle S_{i}\rangle$ and $q_{i}=\langle (S_{i})^{2}\rangle, i=1,...,L)$ perpendicular to the surface of the film corresponding to $L$ parallel distinct layers can be obtained by conventional EFT formulation based on differential operator technique and decoupling approximation (DA) \cite{kaneyoshi_0},
\begin{eqnarray}\label{eq2}
\nonumber
m_{1}&=&[1+m_{1}B_{1}+q_{1}(A_{1}-1)]^{z}[1+m_{2}B_{2}+q_{2}(A_{2}-1)]F_{1}(x)|_{x=0},\\
\nonumber
m_{p}&=&[1+m_{p}B_{2}+q_{p}(A_{2}-1)]^{z}[1+m_{p-1}B_{2}+q_{p-1}(A_{2}-1)],\\
\nonumber
&&+[1+m_{p+1}B_{2}+q_{p+1}(A_{2}-1)]F_{1}(x)|_{x=0}, \\
\nonumber
m_{L}&=&[1+m_{L}B_{1}+q_{L}(A_{1}-1)]^{z}[1+m_{L-1}B_{2}+q_{L-1}(A_{2}-1)]F_{1}(x)|_{x=0},\\
\nonumber
q_{1}&=&[1+m_{1}B_{1}+q_{1}(A_{1}-1)]^{z}[1+m_{2}B_{2}+q_{2}(A_{2}-1)]F_{2}(x)|_{x=0},\\
\nonumber
q_{p}&=&[1+m_{p}B_{2}+q_{p}(A_{2}-1)]^{z}[1+m_{p-1}B_{2}+q_{p-1}(A_{2}-1)],\\
\nonumber
&&+[1+m_{p+1}B_{2}+q_{p+1}(A_{2}-1)]F_{2}(x)|_{x=0}, \\
\nonumber
q_{L}&=&[1+m_{L}B_{1}+q_{L}(A_{1}-1)]^{z}[1+m_{L-1}B_{2}+q_{L-1}(A_{2}-1)]F_{2}(x)|_{x=0},\\
\end{eqnarray}
where $2\leq p\leq L-1$, $z$ is the coordination number of the lattice, and the coefficients $A_{i}$ and $B_{i}$ are defined as  $A_{1}=\cosh(J_{s}\nabla)$, $A_{2}=\cosh(J_{b}\nabla)$, $B_{1}=\sinh(J_{s}\nabla)$ and $B_{2}=\sinh(J_{b}\nabla)$. In the present work, we will focus on the ferromagnetic films in a simple cubic lattice structure, i.e. $z=6$.
The functions $F_{1}(x)$ and $F_{2}(x)$ in Eq. (\ref{eq2}) are then given by
\begin{eqnarray}
\nonumber
F_{1}(x)&=&\frac{2\sinh(\beta x)}{2\cosh(\beta x)+\exp(-\beta D)},\\
\nonumber
F_{2}(x)&=&\frac{2\cosh(\beta x)}{2\cosh(\beta x)+\exp(-\beta D)},
\end{eqnarray}
where $\beta$ is the inverse of the reduced temperature.

With the help of the Binomial expansion, Eq. (\ref{eq2}) can be written as follows:
\begin{eqnarray}\label{eq3}
\nonumber
m_{1}&=&\sum_{i=0}^{z}\sum_{j=0}^{i}\sum_{k=0}^{1}\sum_{l=0}^{k}K_{1}^{(1)}(i,j,k,l)m_{1}^{j}m_{2}^{l}q_{1}^{i-j}q_{2}^{k-l},\\
\nonumber
m_{p}&=&\sum_{i=0}^{z}\sum_{j=0}^{i}\sum_{k=0}^{1}\sum_{l=0}^{k}\sum_{n=0}^{1}\sum_{t=0}^{n}K_{2}^{(1)}(i,j,k,l,n,t)
m_{p-1}^{l}m_{p}^{j}m_{p+1}^{t}q_{p-1}^{k-l}q_{p}^{i-j}q_{p+1}^{n-t},\\
\nonumber
m_{L}&=&\sum_{i=0}^{z}\sum_{j=0}^{i}\sum_{k=0}^{1}\sum_{l=0}^{k}K_{1}^{(1)}(i,j,k,l)m_{L}^{j}m_{L-1}^{l}q_{L}^{i-j}q_{L-1}^{k-l},\\
\nonumber
q_{1}&=&\sum_{i=0}^{z}\sum_{j=0}^{i}\sum_{k=0}^{1}\sum_{l=0}^{k}K_{1}^{(2)}(i,j,k,l)m_{1}^{j}m_{2}^{l}q_{1}^{i-j}q_{2}^{k-l},\\
\nonumber
q_{p}&=&\sum_{i=0}^{z}\sum_{j=0}^{i}\sum_{k=0}^{1}\sum_{l=0}^{k}\sum_{n=0}^{1}\sum_{t=0}^{n}K_{2}^{(2)}(i,j,k,l,n,t)
m_{p-1}^{l}m_{p}^{j}m_{p+1}^{t}q_{p-1}^{k-l}q_{p}^{i-j}q_{p+1}^{n-t},\\
\nonumber
q_{L}&=&\sum_{i=0}^{z}\sum_{j=0}^{i}\sum_{k=0}^{1}\sum_{l=0}^{k}K_{1}^{(2)}(i,j,k,l)m_{L}^{j}m_{L-1}^{l}q_{L}^{i-j}q_{L-1}^{k-l},\\
\end{eqnarray}
with the coefficients
\begin{eqnarray}\label{eq4}
\nonumber
K_{1}^{(\alpha)}(i,j,k,l)&=&
\left(
\begin{array}{c}
z \\
i \\
\end{array}
\right)
\left(
\begin{array}{c}
i \\
j \\
\end{array}
\right)
\sum_{x=0}^{i-j}\sum_{y=0}^{k-l}
\left(
\begin{array}{c}
i-j \\
x \\
\end{array}
\right)
\left(
\begin{array}{c}
k-l \\
y \\
\end{array}
\right)\\
\nonumber
&&\times
(-1)^{i+k-j-l-x-y}\Theta_{\alpha}(x,j,y,l),\\
\nonumber
K_{2}^{(\alpha)}(i,j,k,l,n,t)&=&
\left(
\begin{array}{c}
z \\
i \\
\end{array}
\right)
\left(
\begin{array}{c}
i \\
j \\
\end{array}
\right)
\sum_{x=0}^{i+k+n-j-l-t}
\left(
\begin{array}{c}
i+k+n-j-l-t \\
x \\
\end{array}
\right)\\
&&\times(-1)^{i+k+n-j-l-t-x}\Theta_{\alpha}(0,0,x,j+l+t),
\end{eqnarray}
where
\begin{equation}\label{eq5}
\Theta_{\alpha}(k,m,l,n)=A_{1}^{k}A_{2}^{m}B_{1}^{l}B_{2}^{n}F_{\alpha}(x)|_{x=0},\quad \alpha=1,2.
\end{equation}
Consequently, applying the Binomial expansion in Eq. (\ref{eq5}) yields
\begin{eqnarray}\label{eq6}
\nonumber
\Theta_{\alpha}(k,l,m,n)&=&2^{-(k+l+m+n)}\sum_{r=0}^{k}\sum_{s=0}^{l}\sum_{t=0}^{m}\sum_{v=0}^{n}
\left(
\begin{array}{c}
k \\
r \\
\end{array}
\right)
\left(
\begin{array}{c}
l \\
s \\
\end{array}
\right)
\left(
\begin{array}{c}
m \\
t \\
\end{array}
\right)
\left(
\begin{array}{c}
n \\
v \\
\end{array}
\right)
(-1)^{l+n-s-v}\\
&&\times\exp[(2r+2s-k-l)J_{s}\nabla]
\exp[(2t+2v-m-n)J_{b}\nabla]F_{\alpha}(x)|_{x=0}.
\end{eqnarray}

Using Eq. (\ref{eq6}), the coefficients in Eq. (\ref{eq4}) can be numerically evaluated with the help of the relation $\exp(a\nabla)f(x)=f(x+a)$ for an arbitrary $a$. Once the coefficients in Eq. (\ref{eq6}) are numerically evaluated, we obtain a system of coupled non-linear equations from Eq. (\ref{eq3}) which contains $2L$ unknowns. The longitudinal magnetization $m_{i}$, as well as the quadrupolar moment $q_{i}$ of each layer can be obtained from numerical solution of Eq. (\ref{eq3}). Then, the total magnetization and quadrupolar moment of the entire system can be defined as
\begin{equation}\label{eq7}
m=\frac{1}{L}\sum_{i=1}^{L}m_{i}, \qquad q=\frac{1}{L}\sum_{i=1}^{l}q_{i}.
\end{equation}

Since, the magnetization of the entire system is close to zero in the vicinity of the second order phase transition, the transition temperature can be obtained by linearizing Eq. (\ref{eq3}), i.e.
\begin{eqnarray}\label{eq8}
\nonumber
m_{1}&=&\sum_{i=1}^{z}\sum_{k=0}^{1}K_{1}^{(1)}(i,1,k,0)m_{1}q_{1}^{i-1}q_{2}^{k}+\sum_{i=0}^{z}K_{1}^{(1)}(i,0,1,1)m_{2}q_{1}^{i},\\
\nonumber
m_{p}&=&\sum_{i=0}^{z}\sum_{n=0}^{1}K_{2}^{(1)}(i,0,1,1,n,0)m_{p-1}q_{p}^{i}q_{p+1}^{n}
+\sum_{i=1}^{z}\sum_{k=0}^{1}\sum_{n=0}^{1}K_{2}^{(1)}(i,1,k,0,n,0)m_{p}q_{p-1}^{k}q_{p}^{i-1}q_{p+1}^{n}\\
\nonumber
&&+\sum_{i=0}^{z}\sum_{k=0}^{1}K_{2}^{(1)}(i,0,k,0,1,1)m_{p+1}q_{p-1}^{k}q_{p}^{i},\\
\nonumber
m_{L}&=&\sum_{i=1}^{z}\sum_{k=0}^{1}K_{1}^{(1)}(i,1,k,0)m_{L}q_{L}^{i-1}q_{L-1}^{k}+\sum_{i=0}^{z}K_{1}^{(1)}(i,0,1,1)m_{L-1}q_{L}^{i},\\
\nonumber
q_{1}&=&\sum_{i=0}^{z}\sum_{k=0}^{1}K_{1}^{(2)}(i,0,k,0)q_{1}^{i}q_{2}^{k},\\
\nonumber
q_{p}&=&\sum_{i=0}^{z}\sum_{k=0}^{1}\sum_{n=0}^{1}K_{2}^{(2)}(i,0,k,0,n,0)q_{p-1}^{k}q_{p}^{i}q_{p+1}^{n}\\
q_{L}&=&\sum_{i=0}^{z}\sum_{k=0}^{1}K_{1}^{(2)}(i,0,k,0)q_{L}^{i}q_{L-1}^{k}.
\end{eqnarray}

Critical temperature as a function of the system parameters can be determined from $\mathrm{det}(A)=0$ where $A$ is the coefficients matrix of the set of $2L$ linear equations in Eq. (\ref{eq8}). We note that the determination of the transition temperature should be treated carefully since as it was previously stated in Ref. \cite{sarmento}, from the many formal solutions of $\mathrm{det}(A)=0$, we have to choose the one corresponding to the highest possible transition temperature.

According to the finite-size scaling theory \cite{barber}, the deviation of the thickness dependent critical temperature $T_{c}(L)$ of a thin ferromagnetic film from the bulk critical temperature $T_{c}(\infty)$ can be measured in terms of a scaling relation
\begin{equation}\label{eq9}
\varepsilon=1-T_{c}(L)/T_{c}(\infty)\propto L^{-\lambda},
\end{equation}
for sufficiently thicker films where $\lambda$ is called the shift exponent which is related to the correlation length exponent of the bulk system as $\lambda=1/\nu_{b}$. The exponent $\lambda$ can be extracted from numerical data by plotting $\varepsilon$ versus $L$ curves for sufficiently thick films in a $\mathrm{log-log}$ scale then fitting the resultant curve using the standard linear regression method.

\section{Results and Discussion}\label{results}
In this section, we will discuss the effect of the presence of a surface, as well as the single-ion anisotropy on the universality features of layering transition characteristics of the system. At this point, we note that the value of the bulk exchange interaction $J_{b}$ is fixed to unity, and we also use the normalized surface to bulk ratio of exchange interactions $R=J_{s}/J_{b}$, as well as the reduced single-ion anisotropy energy $\Delta=D/J_{b}$ and reduced temperature $k_{B}T/J_{b}$ throughout the calculations. For simplicity, the exchange couplings are restricted to the ferromagnetic case.

\begin{figure}[!h]
\includegraphics[width=8cm]{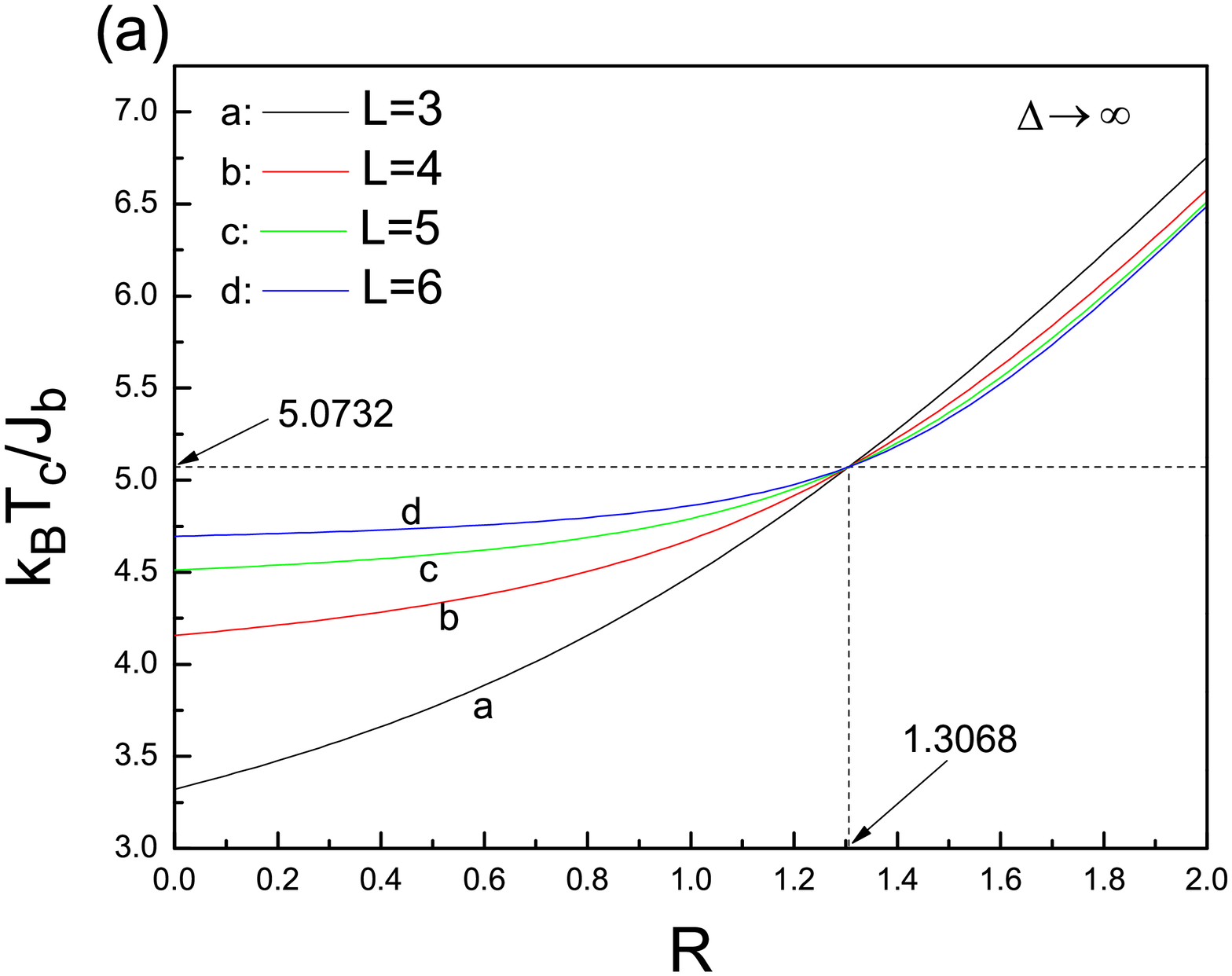}
\includegraphics[width=8.3cm]{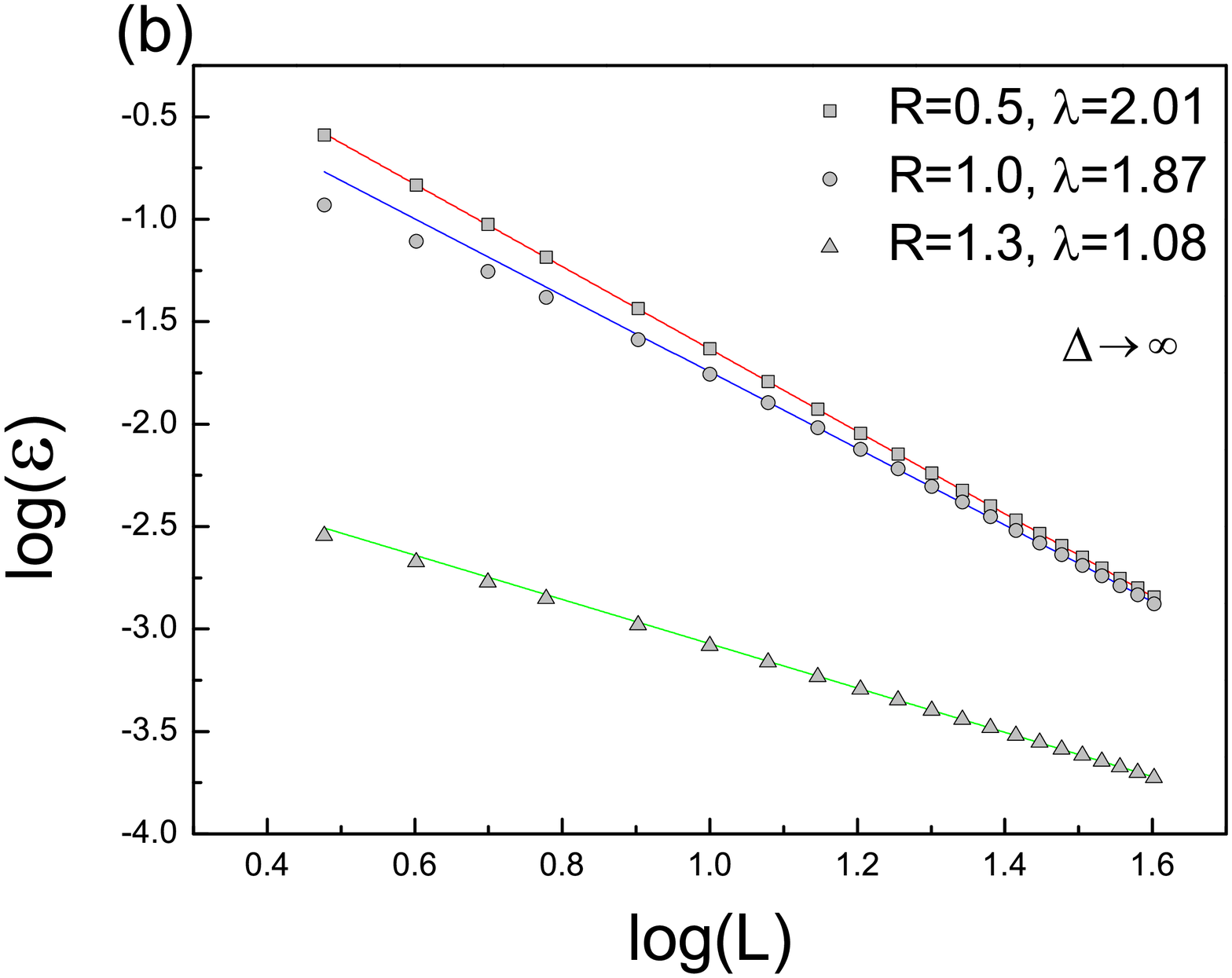}\\
\caption{(Color online) (a) Phase diagrams of Blume-Capel thin film in a $(k_{B}T_{c}/J_{b}-R)$ plane corresponding to the highly anisotropic limit $\Delta\rightarrow\infty$ for various film thickness $L$. (b) Variation of the shift exponent $\lambda$ with surface to bulk ratio of exchange couplings $R$ corresponding to $\Delta\rightarrow\infty$ limit.}\label{fig1}
\end{figure}
In order to provide a testing ground for our calculations, we have primarily studied the phase diagrams in a $(k_{B}T_{c}/J_{b}-R)$ plane corresponding to a highly anisotropic case $\Delta\rightarrow\infty$ with various film thickness $L$ in Fig. \ref{fig1}(a). It is a well known fact that the behavior of the system reduces to that of a spin-1/2 thin film in this limit. Namely, the special value of the surface to bulk ratio of exchange interactions at which the critical temperature is independent of thickness $L$ is found as $R_{c}=1.3068$, and the corresponding bulk transition temperature is $k_{B}T_{c}/J_{b}=5.0732$. These results are identical to those obtained in previous works using EFT \cite{kaneyoshi_1,sarmento}. It is well established that for $R<R_{c}$, the bulk dominates  against the surface whereas for $R>R_{c}$, the surface effects become prominent. In order to examine the universality behavior in this limiting case (i.e. $\Delta\rightarrow\infty$), we have extracted the shift exponent $\lambda$ by computing the transition temperature $k_{B}T_{c}/J_{b}$ as a function of thickness $L$ according to Eq. (\ref{eq9}). In order to precisely cover the critical region, the obtained data have been fitted for those providing the condition $0.01\leq \varepsilon \leq 0.1$ \cite{binder_thin} which requires to consider the transition temperatures of the films for $L\geq10$ in fitting procedure. The results are depicted in Fig. \ref{fig1}(b). For $R=0.5$, at which the bulk highly dominates against the surface, the estimated value of the shift exponent is $\lambda=2.01$ indicating the fact that the system exhibits likely  a three dimensional character. On the other hand, in the vicinity of the special point $R=1.3 \approx R_{c}$, the surface effects become dominant, and we obtain $\lambda=1.08$ which reveals a two dimensional character even for thicker films. We can see from this result that as $R$ increases continuously from zero to $R_{c}$, the system may exhibit a dimensional crossover due to the presence of surface exchange enhancement. In addition, the value $\lambda=1.87$ corresponding to $R=1$ can be compared with the results $\lambda=1.56$ and $\lambda=2.002$ obtained by MC simulations \cite{binder_thin} and MFT \cite{laosiritaworn}, respectively.

\begin{figure*}[!h]
  \includegraphics[width=8cm]{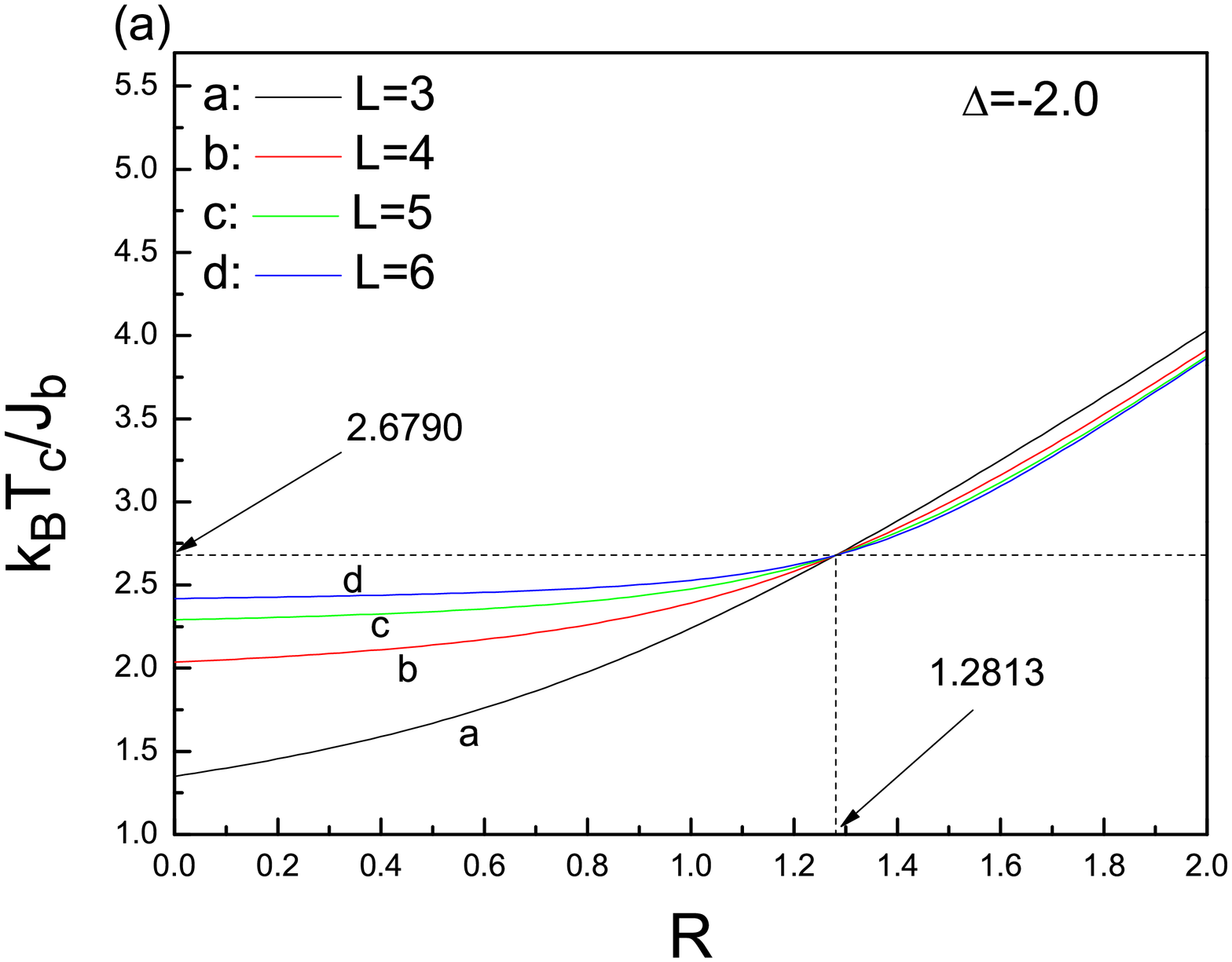}
  \includegraphics[width=8cm]{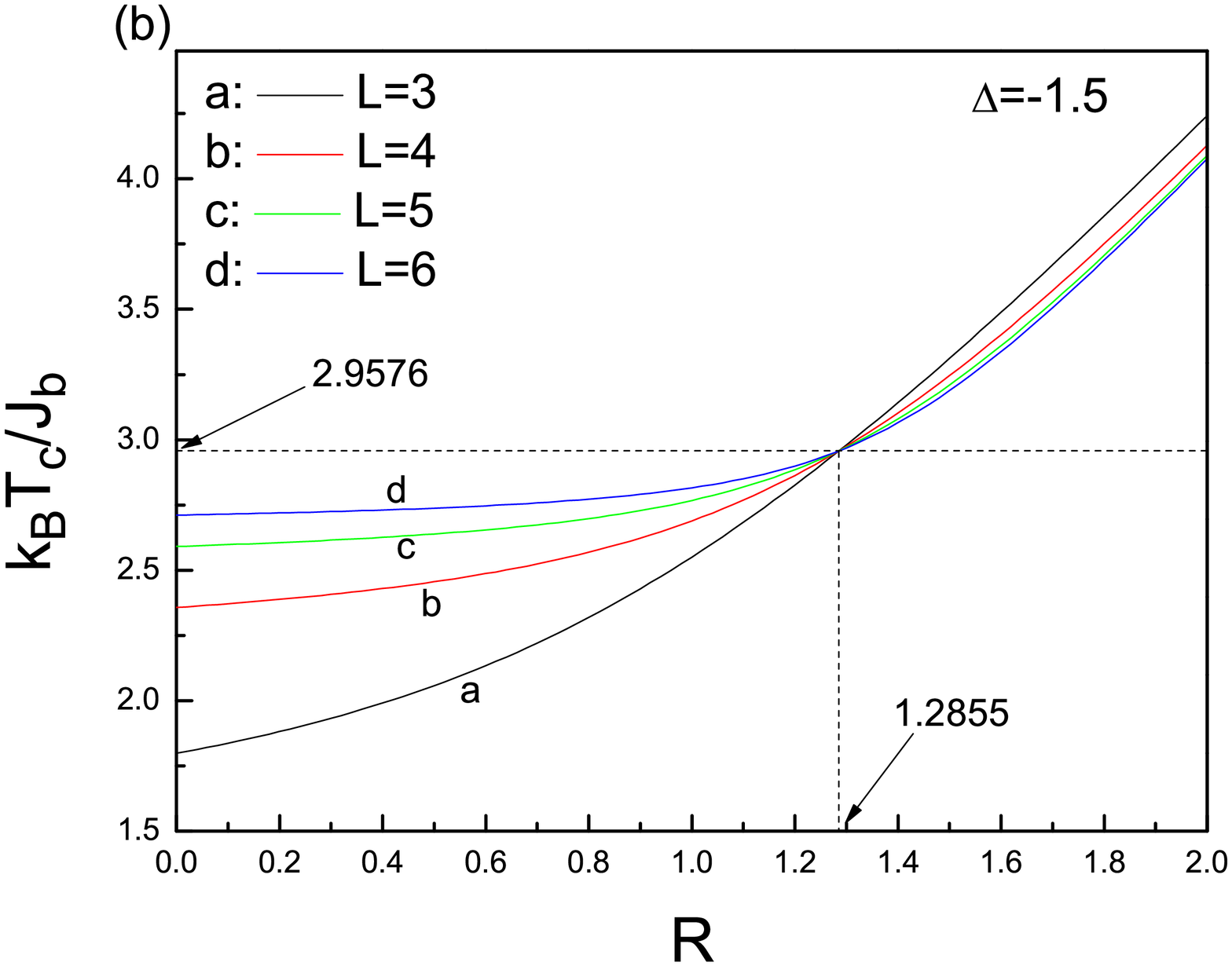}\\
  \includegraphics[width=8cm]{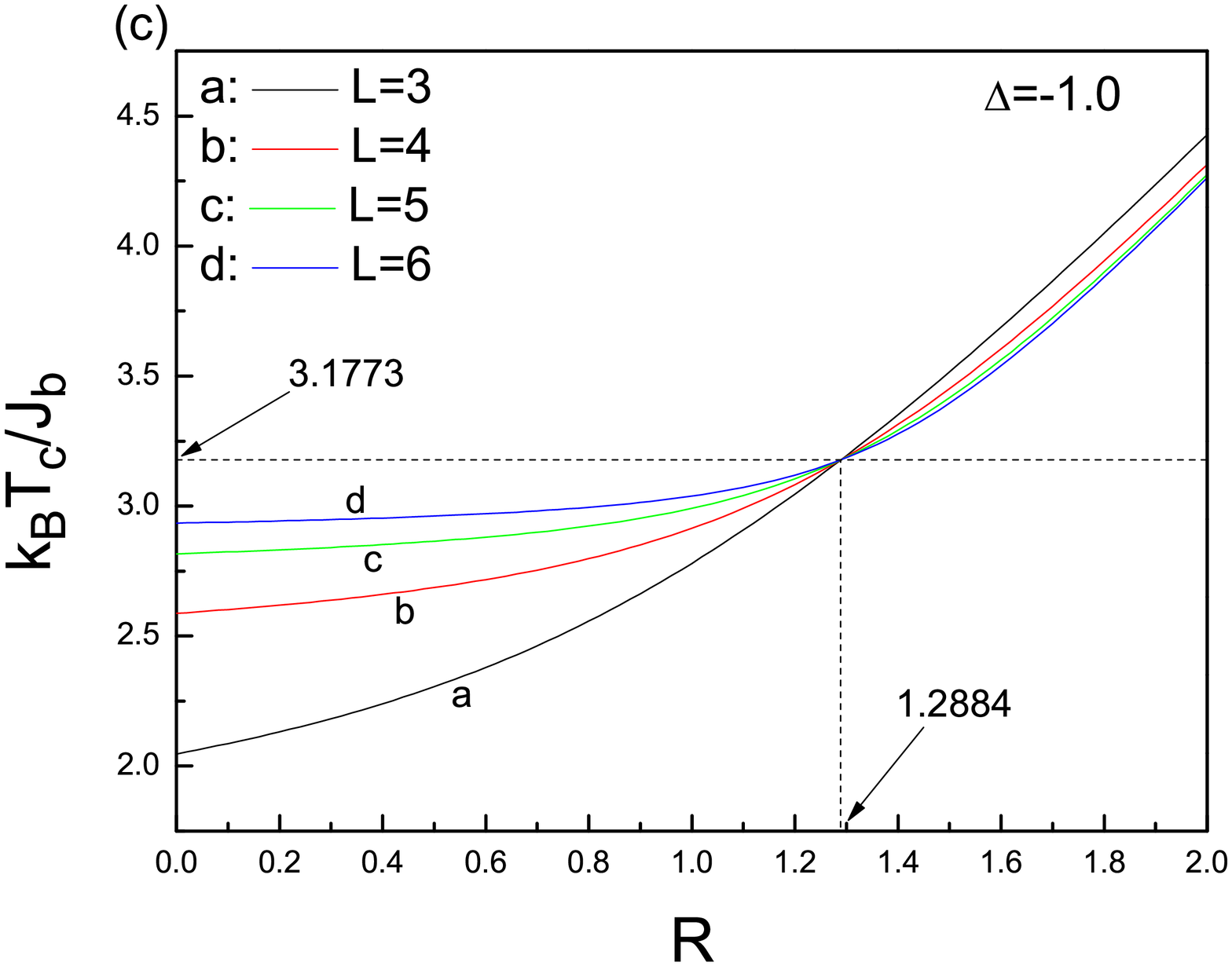}
  \includegraphics[width=8cm]{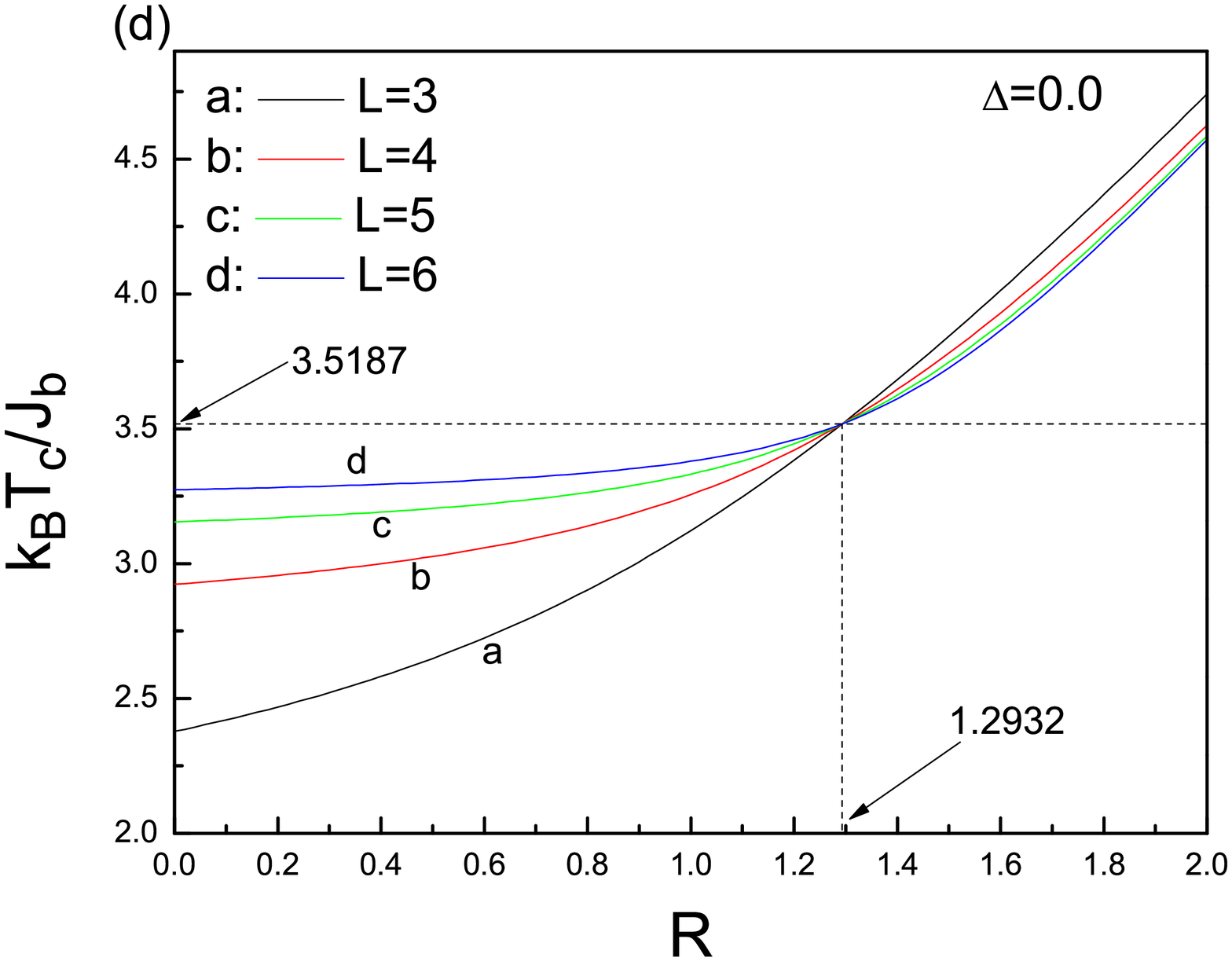}\\
    \caption{(Color online) Phase diagrams of Blume-Capel thin film in a $(k_{B}T_{c}/J_{b}-R)$ plane for various film thickness $L$ corresponding to (a) $\Delta=-2.0$, (b) $\Delta=-1.5$, (c) $\Delta=-1.0$, and (d) $\Delta=0.0$}\label{fig2}
\end{figure*}

In Ref. \cite{tucker}, surface exchange enhancement effects on ultrathin spin-1 Ising films have been examined, and based on the results obtained within MC simulations, it has been concluded that $R_{c}$ is a spin dependent parameter which has a value lower than that obtained for a spin-1/2 system \cite{landau_0}. On the contrary, Ref. \cite{zaim} suggests that $R_{c}$ was independent of crystal field interactions which gives rise to a controversial situation. To the best of our knowledge, this case has not yet been examined in the literature. Therefore, in Fig. \ref{fig2}, we depict the phase diagrams in a $(k_{B}T_{c}/J_{b}-R)$ plane with various thickness $L$ corresponding to some selected values of crystal field interactions $\Delta$. It is clear from Fig. \ref{fig2} that $R_{c}$ has its minimum value for a sufficiently negative crystal field value. For $\Delta=0.0$, we have $R_{c}=1.2932$ which can be compared with $R_{c}=1.45$ of MC simulations \cite{tucker} whereas MFT \cite{aguilera} predicts a spin independent value.

\begin{figure*}[!h]
\includegraphics[width=8cm]{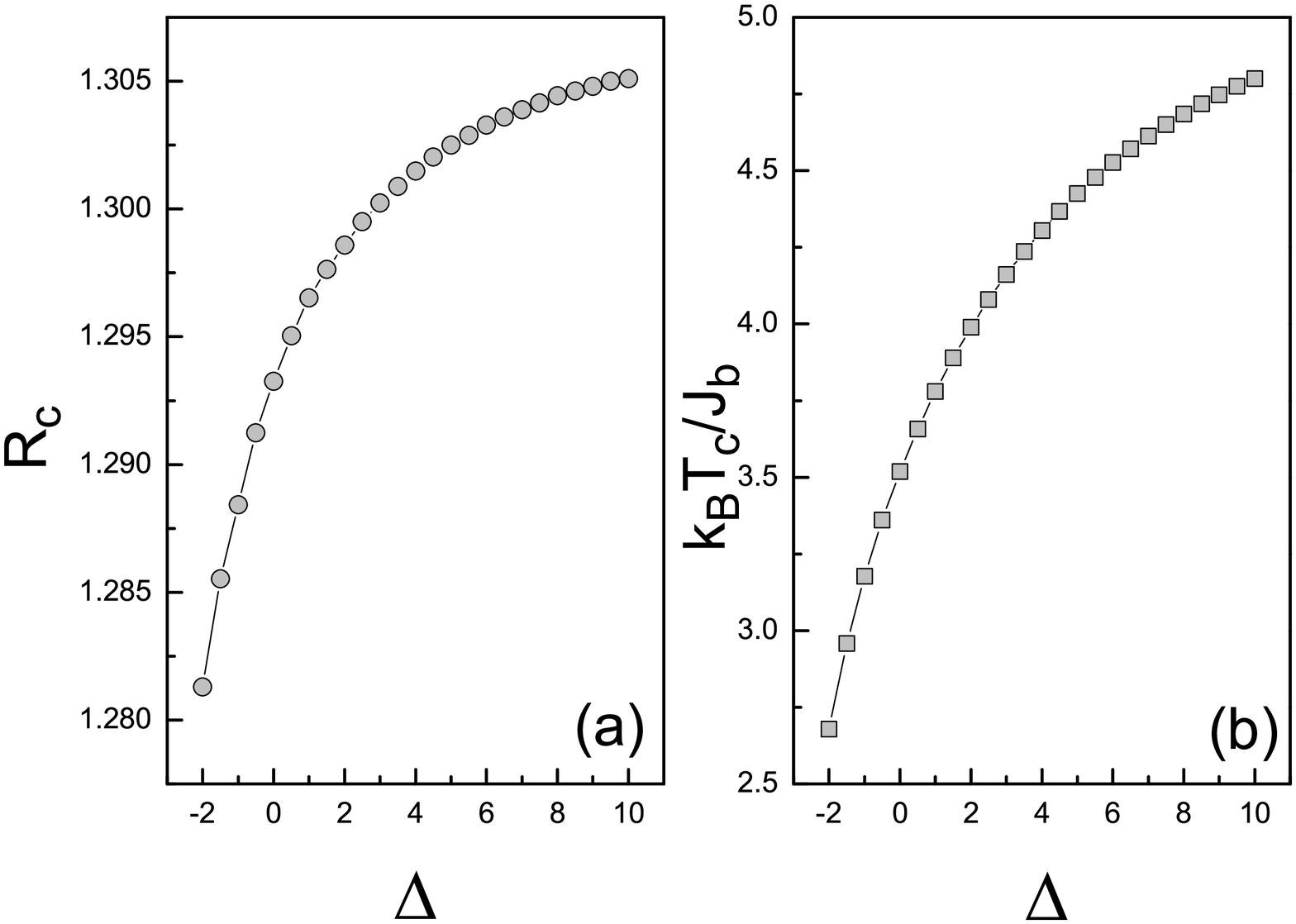}\\
\caption{Dependence of (a) the special point $R_{c}$, (b) corresponding bulk transition temperature $k_{B}T_{c}/J_{b}$ as a function of crystal field interaction $\Delta$.}\label{fig3}
\end{figure*}

Dependencies of the special point $R_{c}$ and corresponding bulk transition temperature $k_{B}T_{c}/J_{b}$ on crystal field interaction $\Delta$ have been clarified and presented in Fig. \ref{fig3}. As seen from Figs. \ref{fig3}(a) and \ref{fig3}(b) that $R_{c}$ and $k_{B}T_{c}/J_{b}$ increase from their minimum values $R_{c}=1.2813$ and $k_{B}T_{c}/J_{b}=2.679$ at $\Delta=-2.0$ and reach to $R_{c}=1.2932$ and $k_{B}T_{c}/J_{b}=3.5187$ at $\Delta=0.0$ and approaches to its spin-1/2 limit $R_{c}=1.3068$ and $k_{B}T_{c}/J_{b}=5.073$ in the limit $\Delta\rightarrow\infty$. A similar investigation was made on thin films consisting of a spin-1/2 bulk coated with a spin-3/2 surface with using EFT \cite{kaneyoshi_balcerzak}, and it has been reported that $R_{c}$ has its maximum value at sufficiently negative crystal fields, and decreases continuously with increasing $\Delta$ in positive direction.

\begin{figure*}[!h]
\includegraphics[width=8cm]{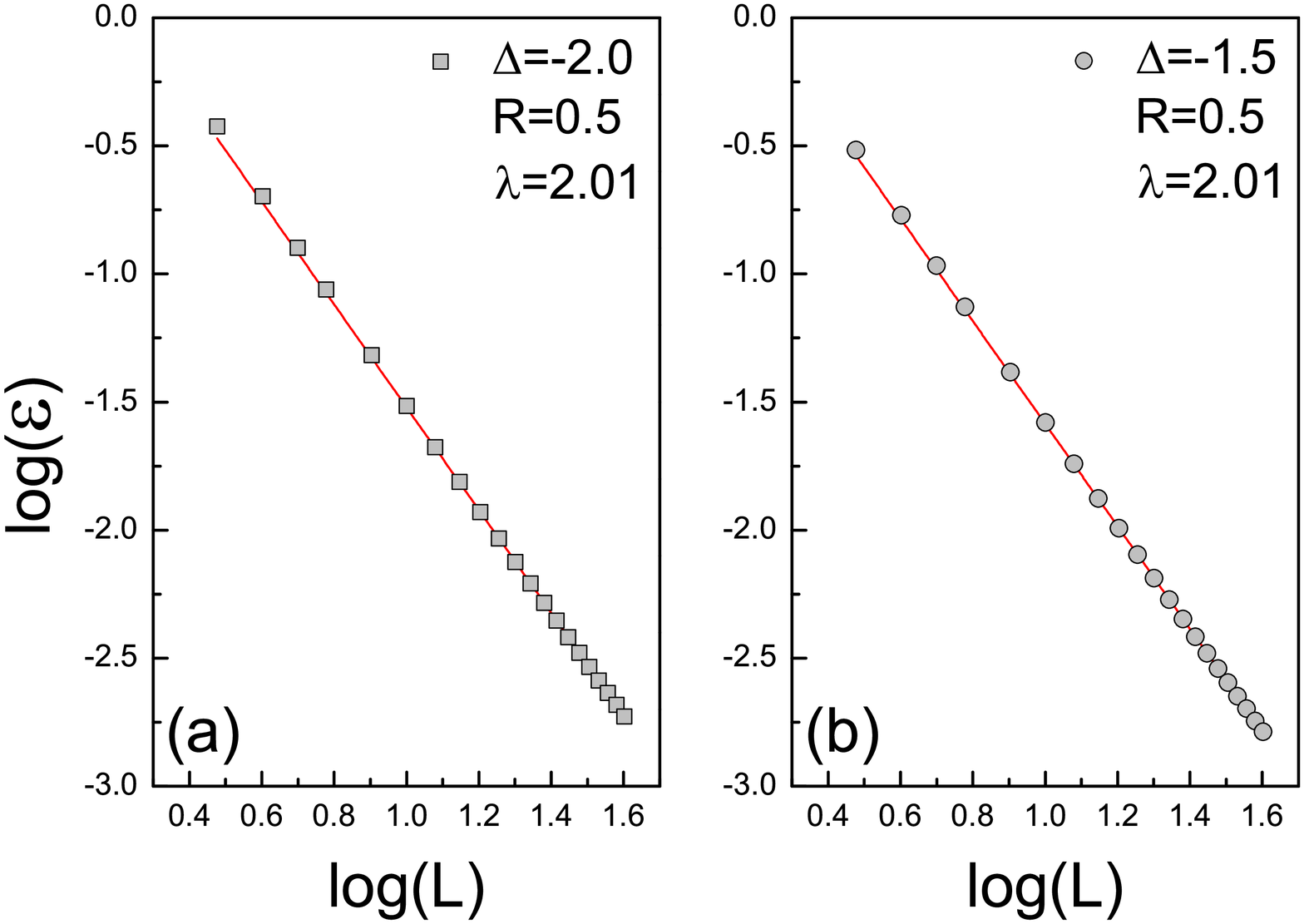}
\includegraphics[width=8cm]{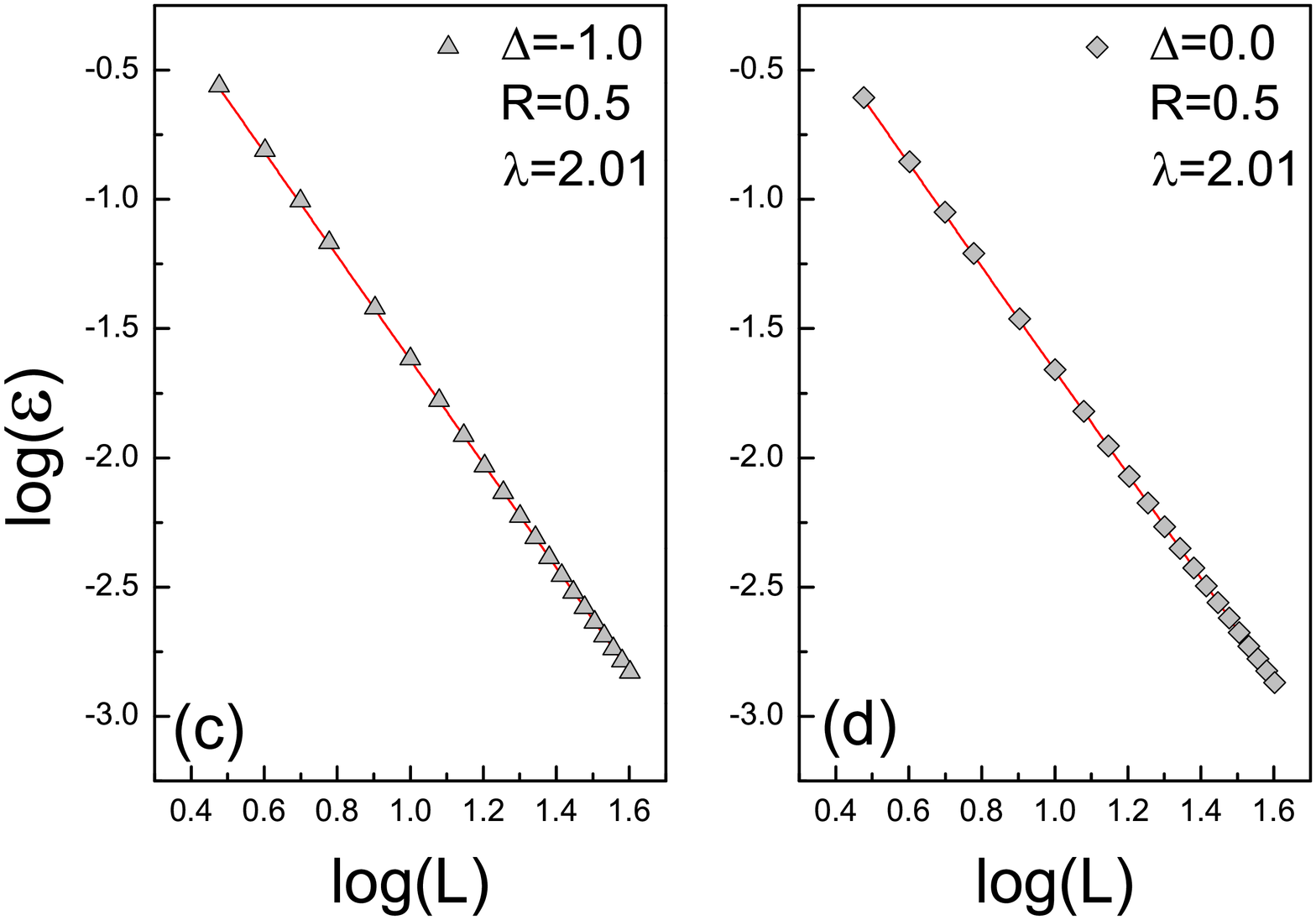}\\
\caption{(Color online) Variation of the shift exponent $\lambda$ for weak surface coupling $R=0.5$ corresponding to some selected values of crystal field interaction (a) $\Delta=-2.0$, (b) $\Delta=-1.5$, (c) $\Delta=-1.0$ and (d) $\Delta=0.0$. }\label{fig4}
\end{figure*}
\begin{figure*}[!h]
\includegraphics[width=8cm]{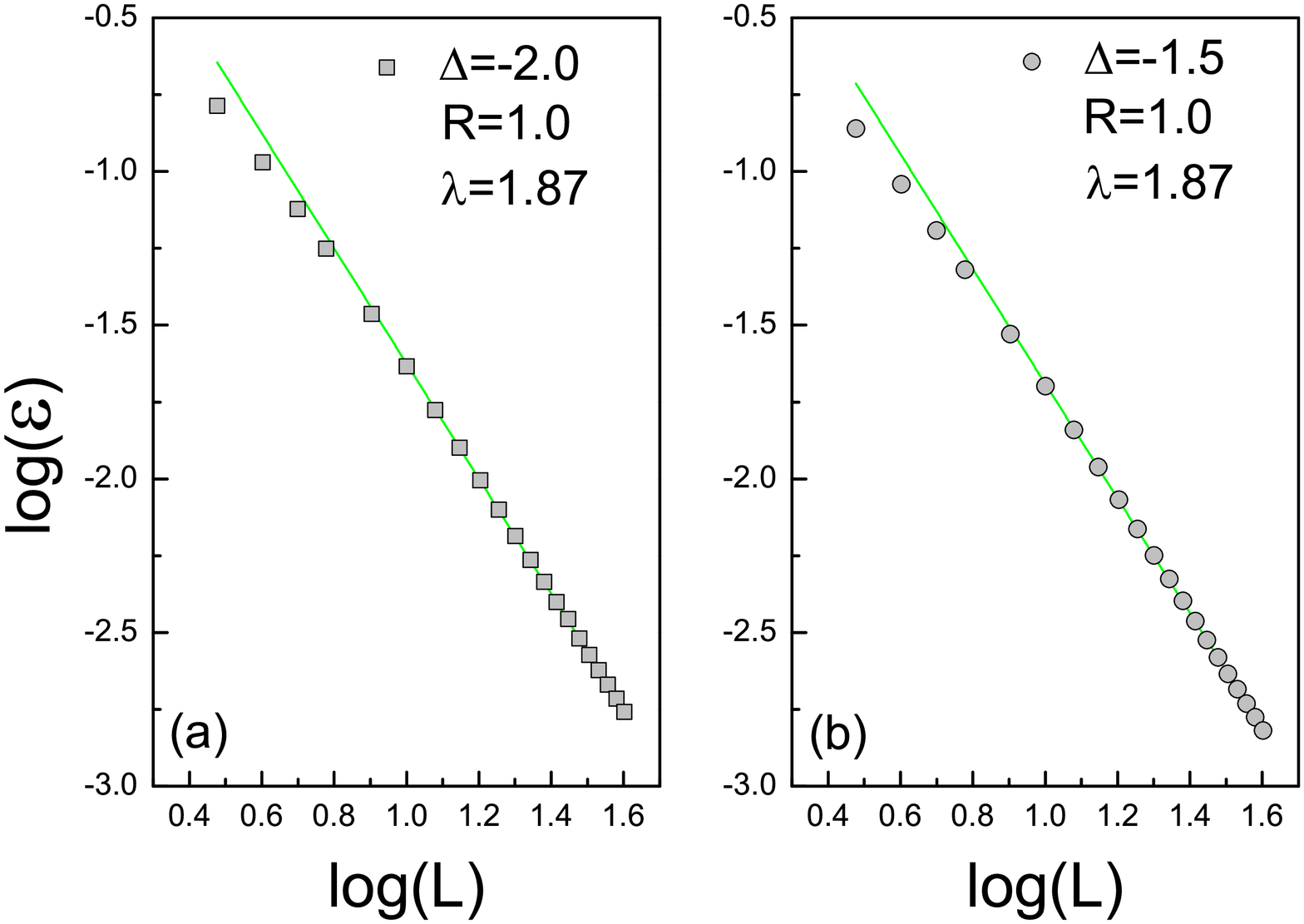}
\includegraphics[width=8cm]{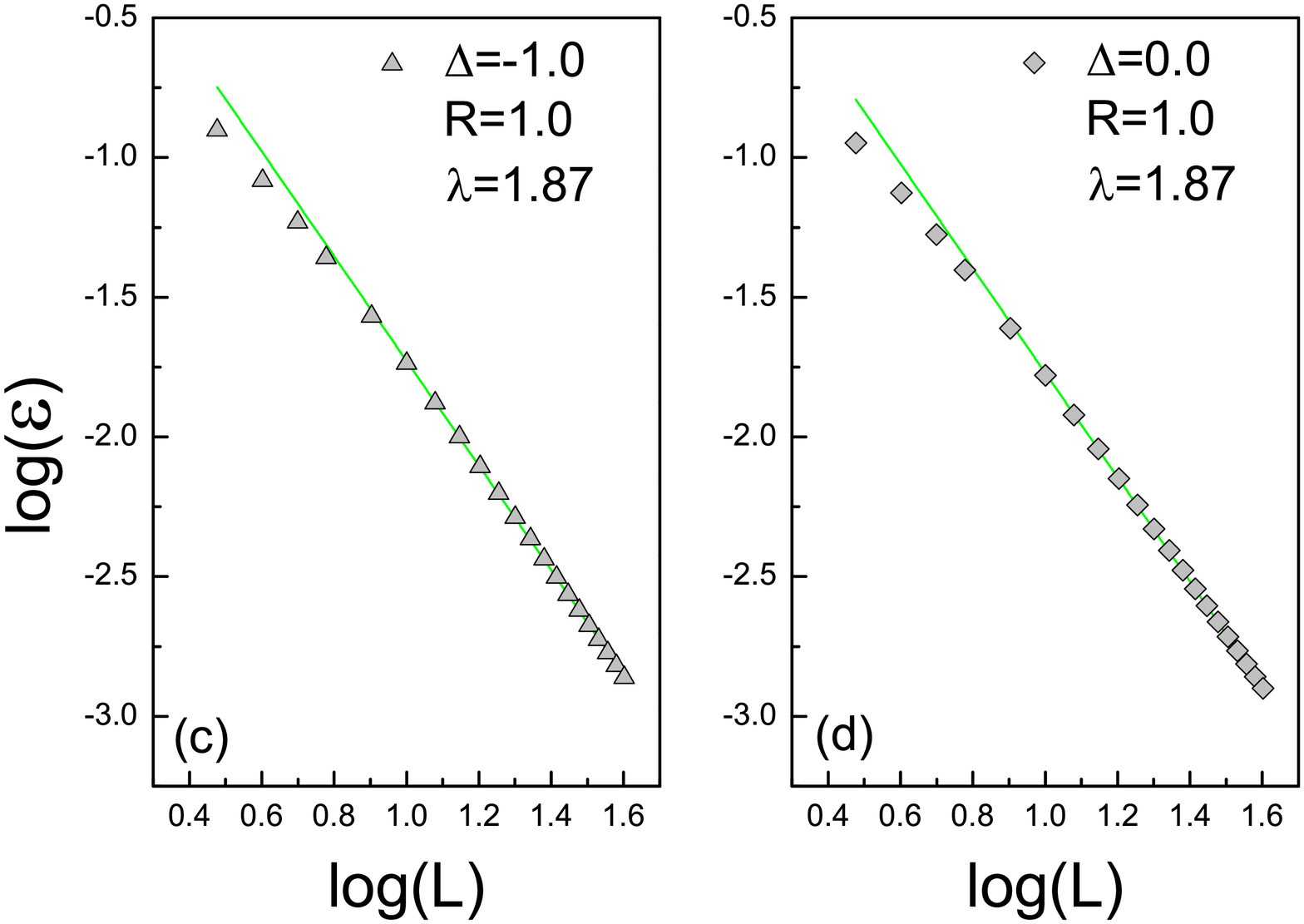}\\
\caption{(Color online) Variation of the shift exponent $\lambda$ for moderate surface coupling $R=1.0$ corresponding to some selected values of crystal field interaction (a) $\Delta=-2.0$, (b) $\Delta=-1.5$, (c) $\Delta=-1.0$ and (d) $\Delta=0.0$.}\label{fig5}
\end{figure*}

Finally, in Figs. \ref{fig4} and \ref{fig5}, the effect of surface exchange interactions and the presence of crystal field interactions on the universality behavior of thin film system have been investigated by examining the variation of the shift exponent with $\Delta$. Only the transition temperatures corresponding to thickness values $L$ $(L\geq10)$ which obey the criteria $0.01\leq \varepsilon\leq0.1$ are taken into account in the linear fitting process. In Fig. \ref{fig4}, a weak surface coupling such as $R=0.5$ has been considered. In this case the exponent value is estimated as $\lambda=2.01$  whereas for a moderate ratio of surface to bulk exchange interactions such as $R=1.0$, we have $\lambda=1.87$ which are identical to the values obtained for spin-1/2 system in Fig. \ref{fig1}(b). Hence, we can conclude that the shift exponent $\lambda$ is found to be independent of $\Delta$.

\section{Conclusions}\label{conclude}
In conclusion, with using EFT, we have studied the universal behavior and critical phenomena in a ferromagnetic thin film described by a spin-1 Blume-Capel Hamiltonian. It has been shown that crystal field interaction $\Delta$ plays an important role in layering transitions. Namely, we have found that the critical value of surface to bulk ratio of exchange interactions $R_{c}$ strictly depends on the crystal field interactions $\Delta$. Strictly speaking, for sufficiently negative values of $\Delta$, $R_{c}$ exhibits a minimum, and continuously increases as $\Delta$ increases. In the highly anisotropic limit $\Delta\rightarrow\infty$, it approaches to the value $R_{c}=1.3068$ which corresponds to the critical coupling of spin-1/2 thin film system.  In this context, we believe that recent calculations reported in the literature should be treated carefully.

Apart from these, dependence of universality on the surface exchange enhancement, as well as the crystal field interactions has been clarified by examining the shift exponent $\lambda$ for a wide range of film thickness, $3\leq L \leq40$. In the presence of surface exchange enhancement, the exponent $\lambda$ approaches to unity, being independent of crystal fields. In this regard, we have concluded that a dimensional crossover may originate as the surface becomes dominant against the bulk, and in terms of the exponent $\lambda$, a ferromagnetic spin-1/2 thin film is in the same universality class with its spin-1 counterpart.




\end{document}